\documentstyle[11pt]{article}
\begin{document}

\centerline{\Large\bf
Variational Monte Carlo Studies of $t$-$J$-type Models
}
\par\bigskip\bigskip

\centerline{
Hisatoshi {\sc Yokoyama}
\footnote{E-mail: yoko@cmpt01.phys.tohoku.ac.jp},
Yoshio {\sc Kuramoto}
\footnote{E-mail: kuramoto@cmpt01.phys.tohoku.ac.jp} and
Masao {\sc Ogata}$^\dagger$
\footnote{E-mail: ogata@sola.c.u-tokyo.ac.jp}
}
\par\bigskip\bigskip

\centerline{\it
Department of Physics, Tohoku University,
Sendai 980
}
\centerline{\it
$^\dagger$Institute of Physics, College of Arts and Sciences,
University of Tokyo, Tokyo 153.
}
\par\bigskip

\centerline{(Received January 24, 1995)}


\begin{abstract}
We give a brief review of recent developments by the variational
Monte Carlo method, in addition to some new results.
We discuss $t$-$J$-type models: the ordinary $t$-$J$ model
in one and two dimensions, and the one-dimensional
supersymmetric $t$-$J$ model with the long-range coupling of
inverse square.
\end{abstract}
\baselineskip 14pt

\section{Introduction}
The variational Monte Carlo (VMC) method has been extensively
developed for many-body systems, since the first successful
applications to bosons~\cite{McMillan}and fermions.~\cite{Ceperley}
Generally speaking, it was not until the VMC method appeared
that the many-body variation theory came to give reliable results
without uncontrollable approximations.
And now the VMC is a standard method to estimate expectation
values in this theory.
\par
In solid state physics, the VMC method was first applied to
lattice systems such as the Hubbard model~\cite{HK,GJR,YS1}
and the periodic Anderson model,~\cite{Shiba} and modified
some prior results obtained in the additional approximations.
Since reviews of the recent VMC studies in this field
have already been given,~\cite{ShibaR,OgataR}
we avoid the overlap of the material,
and concentrate on the recent developments for
the ordinary $t$-$J$ model~\cite{AZR} and the long-range
$t$-$J$ model.~\cite{CE}
\par
Before going into specific models, we give a brief
summary of the VMC method in \S2.
In \S3.1 we discuss the one-dimensional (1D) $t$-$J$
model, for which many exact behaviors are known.
Nevertheless, we show how further knowledge is added
by comparing the VMC results with the exact and
other results.
This is ascribed to the merit of the explicit form of
the wave function.
In \S3.2 we discuss the $t$-$J$ model on the square
lattice, comparing with the properties of the 1D system.
In \S4 we develop a new usage of the VMC method,
namely identification of the eigenstates by checking
the variance.
We apply it to the supersymmetric long-range $t$-$J$
model with the coupling of inverse square.
\par

\section{Formulation}
In this section, we summarize the fundamentals of the
VMC method.~\cite{note1}
This method is a combination of the variation theory
with the Monte Carlo technique for evaluating
expectation values.
Following the variation theory, one has to estimate
the energy expectation value with respect to a
given trial function $\psi_\gamma({\bf R})$~\cite{note15}
as a function of $\gamma$.~\cite{note2}
Here $\gamma$ denotes a set of variational parameters
and {\bf R} represents a certain electron configuration.
Next, by using the optimal $\gamma$, one
should calculate physical quantities.
Actually, the variational expectation value of an
operator $O$ is written as,
\begin{equation}
\langle O\rangle_\gamma\!
=\!\frac{\langle\psi_\gamma({\bf R})|O|\psi_\gamma({\bf R})\rangle}
{\langle\psi_\gamma({\bf R})|\psi_\gamma({\bf R})\rangle}
\!=\!\sum_{\bf R}P_\gamma({\bf R})
\frac{O\psi_\gamma({\bf R})}{\psi_\gamma({\bf R})},
\end{equation}
where $P_\gamma({\bf R})$ is the probability density
for a configuration {\bf R}:
\begin{equation}
P_\gamma({\bf R})=\frac{|\psi_\gamma({\bf R})|^2}
{\sum_{\bf R'}|\psi_\gamma({\bf R'})|^2}.
\end{equation}
\par
According to the Monte Carlo method, the summation over
{\bf R} is replaced by the importance sampling with weight
$P_\gamma({\bf R})$.
Then, eq.(2.1) is reduced to
\begin{equation}
\langle O\rangle_\gamma\cong\frac{1}{M}
\sum_{m=1}^M\frac{O\psi_\gamma({\bf R}_m)}{\psi_\gamma({\bf R}_m)},
\end{equation}
where ${\bf R}_m$ is the $m$-th sample picked out
and $M$ is the total number of samples.
Since $P_\gamma({\bf R})\ge 0$, there is no sign problem.
In the ordinary usage of the VMC method, one should reduce
the statistical errors by taking sufficient samples
and keeping statistical independence among samples.
\par
Inversely one can take advantage of this statistical
property to identify the eigenstates,~\cite{Ceperley}
as will be seen in \S4.
When one applies the Hamiltonian to
$\psi_\gamma({\bf R})$, the following formula
holds:
\begin{equation}
{\cal H}\psi_\gamma({\bf R})=E_\gamma\psi_\gamma({\bf R})
+\xi_\gamma({\bf R}),
\end{equation}
where $\xi_\gamma({\bf R})$ is orthogonal to
$\psi_\gamma({\bf R})$ and $E_\gamma=\langle{\cal H}\rangle_\gamma$.
If $\psi_\gamma({\bf R})$ is an exact eigenstate,
eq.(2.4) has to be reduced to the Schr\"odinger equation,
so that $\xi_\gamma({\bf R})=0$.
This relation and the definition of variance,
\begin{equation}
\sigma_E^2=\langle({\cal H}-E_\gamma)^2\rangle_\gamma
=\frac{\langle\xi_\gamma({\bf R})|\xi_\gamma({\bf R})\rangle}
{\langle\psi_\gamma({\bf R})|\psi_\gamma({\bf R})\rangle}
\end{equation}
lead to the fact that there is no statistical fluctuation
for an exact eigenstate.~\cite{note3}
Then, $E_\gamma$ becomes the eigenenergy.
On the other hand, an approximate function results
in an appreciable variance in proportion to $1/\sqrt{M}$.
Thus, in the actual VMC computations, the vanishing of the
fluctuation can be definitely distinguished.
\par
Now let us turn to the trial wave function.
For $t$-$J$-type models two-body Gutzwiller-Jastrow-type
wave functions,
\begin{equation}
\Psi=\prod_{j\ell}\prod_{\sigma\sigma'}
\left[1-\left(1-\eta(r_{j\ell})\right)n_{j\sigma}n_{\ell\sigma'}\right]\Phi
\end{equation}
are often used, where $\eta(r_{j\ell})$ is a spin-independent
correlation factor ($\eta(0)=0$) with
$r_{j\ell}=|{\bf r}_j-{\bf r}_\ell|$,
and $\Phi$ a Hartree-Fock-type wave function.
By specifying $\Phi$ (one-body state) and $\eta$ (two-body effect),
one can describe a variety of states from metallic to ordered or
insulating ones by eq.(2.6).
\par

Leaving individual forms in the following sections,
here we discuss only the Gutzwiller wave function
(GWF)~\cite{Gutzwiller}
$\Psi_{\rm G}$, which is the exact ground state for
the long-range $t$-$J$ model, and is often used as a
starting trial function for its simplicity.
For the GWF, $\eta(r)=1$ and $\Phi=\Phi_{\rm F}$, which
is the Fermi sea.
The properties of the GWF were investigated for the
Hubbard model;~\cite{YS1,YS3,VMG}
the GWF is generally metallic, having a discontinuity
of momentum distribution $n(k)$ at $k=k_{\rm F}$.
And a repulsive intersite correlation factor is needed
to represent the Hubbard model properly.~\cite{YS3}
\par

\section{$t$-$J$ Model in One and Two Dimensions}
In this section we discuss the $t$-$J$ model defined
as
\begin{equation}
{\cal H}={\cal H}_t+{\cal H}_J,
\end{equation}
\begin{equation}
     {\cal H}_t=-t\sum_{\langle ij\rangle\sigma}
     (c_{i\sigma}^\dagger c_{j\sigma}+c_{j\sigma}^\dagger c_{i\sigma}),
\end{equation}
\begin{equation}
     {\cal H}_J=J\sum_{\langle ij\rangle}
     ({\bf S}_i\cdot {\bf S}_{j}-\frac{1}{4}n_in_{j}),
\end{equation}
in the subspace with no double occupancy of each site
with $t,J\geq0$, and $\langle ij\rangle$ denotes a
nearest-neighbor pair.
Spin operators vanish when they act on empty sites.
Henceforth we take $t$ as the unit of energy.

\par
When we regard eq.(3.1) as an effective Hamiltonian of
the Hubbard model or the d-p model in the strong coupling
regime, the following term of the second-order perturbation
has to be considered:
\begin{equation}
{\cal H}_3 = -\frac{J_3}{4}\sum_{j,\tau\ne\tau',\sigma}
(c^\dagger_{j-\sigma}c_{j-\sigma}
c^\dagger_{j+\tau\sigma}c_{j+\tau'\sigma}
+c^\dagger_{j+\tau-\sigma}c_{j-\sigma}
c^\dagger_{j\sigma}c_{j+\tau'\sigma}) ,
\end{equation}
where $\tau$ and $\tau'$ indicate vectors to the
nearest-neighbor sites.
When eq.(3.4) is derived from the Hubbard model,
$J_3=J$,~\cite{Hirsch,GJR} while $J_3/J<0$ (ferromagnetic)
when derived from the d-p model with plausible
parameters for high-$T_{\rm c}$ superconductors.~\cite{Matsukawa}
\par\medskip\noindent
{\it 3.1\quad One-dimensional case}
\par
Many properties in the one-dimensional (1D) model of eq.(3.1)
have already been elucidated by various exact methods.~\cite{exact}
In the following, we would like to show what kind of new
knowledge the variation theories have added to those.
\par
\marginpar{\fbox{Fig. 1}}

First of all, we take the GWF for this system.
In Fig.1 we show the total energy for the supersymmetric case
($J/t=2$) as a function of electron density $n=N_{\rm e}/N$
($N_{\rm e}$ and $N$ being electron and site number
respectively).~\cite{YO}
Astonishing facts are that the results of the Bethe Ansatz
and the GWF agree almost perfectly, and the non-interacting
system joins for small $n$.
The three results coincide in the low density
limit up to the order of $n^3$, and the difference is
about 0.15\% for $n=1$ between the GWF and the exact
result.~\cite{GJR,VMG}
This good agreement of the two results is
not restricted to the energy;
momentum distribution and correlation functions also show
quantitative agreement, except for the long-range
behaviors~\cite{YO,YO2}discussed later.
\par

Intuitively we can understand these results as follows:
for the supersymmetric case, the Hamiltonian eq.(3.1) becomes
a kind of ``free electron" model, because the kinetic term
eq.(3.2), which tends to make electrons apart, balances
with the attractive term eq.(3.3).
The GWF is the very state to express such a situation.
Thus, away from the supersymmetric case, the GWF has
to be modified to include intersite correlations
repulsive or attractive.
\par

The above agreement seems in close connection
to the fact that the ground state of the long-range $t$-$J$
model discussed in \S4 is nothing but the GWF.
\par

According to the exact result, the Tomonaga-Luttinger liquid
(TLL) is realized in the wide range of $n$ and $J/t$.
For the ground state, correlation functions show
power-law behaviors.
Hellberg and Mele~\cite{HM} introduced a trial function $\Psi_{\rm TLL}$
which has an essentially long-range correlation factor,
\begin{equation}
\eta(r)=\left[\frac{N}{\pi}\sin\left(\frac{\pi}{N}r\right)\right]^\nu,
\end{equation}
and $\Phi=\Phi_{\rm F}$ in eq.(2.6).
This factor becomes repulsive or attractive, according as
the parameter $\nu$ is positive or negative, respectively.
Especially for $\nu<-0.5$ $\Psi_{\rm TLL}$ represents a separate
phase.
Correlation functions of this function show power-law behaviors,
and $\nu$
is analytically related to the TLL exponent $K_\rho$ by
$K_\rho=1/(2\nu+1)$.~\cite{KawaH}
Using this relation, we construct a phase diagram shown in
Fig.2, which should be compared with the diagonalization
result.~\cite{Ogata}
$\Psi_{\rm TLL}$ is successful in the region of relatively
large $J/t$.
\par
\marginpar{\fbox{Fig. 2}}

On the other hand, Fermi-liquid-type wave function
$\Psi_{\rm FL}$~\cite{YS3,YO,YO2} with the correlation factors:
\begin{equation}
\eta(r)=\frac{2}{\pi}\arctan\frac{r}{\zeta}\quad {\rm (repulsive)},
\end{equation}
\begin{equation}
\eta(r)=1+\frac{\alpha}{r^\beta}\quad{\rm (attractive)},
\end{equation}
and $\Phi=\Phi_{\rm F}$, does not show the power-law
behaviors.
For instance, there is a finite discontinuity in $n(k)$.
However bulk properties of $\Psi_{\rm FL}$ quantitatively
agree with those of $\Psi_{\rm TLL}$.
The essential distinction between the correlation factors
of $\Psi_{\rm FL}$ and $\Psi_{\rm TLL}$ is in the long-distance
part; the value of $\eta(1)/\eta(\infty)$ is finite
for every parameter set of $\Psi_{\rm FL}$, but not for
the other.
This fact and a numerical experiment,~\cite{YO2} in which we use
hybrid $\eta(r)$'s of $\Psi_{\rm TLL}$ and $\Psi_{\rm FL}$
connected at a certain value of $r$, indicate that the long-distance
part of the correlation factor determines the long-range behavior,
while the short-distance part does the bulk properties like
energy and the amplitude of the correlation functions.
\par

For the regime of $J/t\sim 0$, a wave function of the form
$\Psi=\chi\phi_{\rm SF}$ is desired on the analogy
of the exact eigenfunction for $J/t\rightarrow 0$.~\cite{OS}
Here $\chi$ is the spin part of the wave function
and $\phi_{\rm SF}$ is the spinless fermion state.
Such a spin-charge separation is lacking in $\Psi_{\rm TLL}$
and $\Psi_{\rm FL}$.
Actually, quantitatively reasonable results have been
obtained by the form $\chi\phi_{\rm SF}$.~\cite{OgataR,Kobayashi}
\par

In the region of small $n$ and $J/t>2$, a spin-gap state
is expected.~\cite{Ogata}
We have not confirmed a corresponding state in
$\Psi_{\rm TLL}$ and $\Psi_{\rm FL}$.
On the other hand, Chen and Lee introduced a trial state
for a gas of singlet pairs, and showed that there is
a region where this function is stabler than $\Psi_{\rm TLL}$.~\cite{CL}
Later, the spin-gap state has been found by using Green's
function Monte Carlo method.~\cite{HMGF}
\par
\marginpar{\fbox{Fig. 3}}

Finally we discuss susceptibilities of charge $\chi_{\rm c}$ and
spin $\chi_{\rm s}$, in connection with the Mott transition
near the half-filling.~\cite{YO2}
In Fig.3 we show $\chi_{\rm c}$ calculated with $\Psi_{\rm TLL}$.
In high density region, $\chi_{\rm c}$ for every value
of $J/t$ is divergent as $n\rightarrow 1$.
For $J/t\le 2$ this divergence is fitted as
$\chi_{\rm c}\propto 1/(1-n)$.
This divergence is due to the strong correlation effect,
in comparison with the non-interacting case.
In the meantime, $\chi_{\rm s}$ converges to a finite value
as $n\rightarrow 1$, as long as $J/t>0$.
These behaviors are consistent with the exact results for
$J/t=0$ and 2.~\cite{KawaTJ}
In comparison with these results, the Brinkman-Rice
transion~\cite{BR} shows different behaviors, namely
$\chi_{\rm c}$ remains finite and $\chi_{\rm s}$ diverges as
$U\rightarrow U_{\rm c}\ (n=1)$ or $n\rightarrow 1
\ (\infty>U>U_{\rm c})$.~\cite{Vollhardt}
\par\medskip\noindent
{\it 3.2\quad Case for square lattice}
\par
Although a lot of studies have done in 2D in connection
with the high-$T_{\rm c}$ superconductivity, there are
not so many definite results.
It is likely in 2D that ordered phases exist
and the phase separation spreads to the region of relatively
small $J/t$ and $n\sim 1$,~\cite{Emery,Puttika} etc.

\par
First we discuss briefly the half-filled
case (the Heisenberg antiferromagnet), where the ground
state has an antiferromagnetic (AF) order.~\cite{Trivedi}
It was also shown that there exist non-AF states which have
very close energy to the ground state by using RVB-type
wave functions,~\cite{Liang,Miyazaki}
However, this type of functions have not yet applied to
the case $n<1$.
\par

\marginpar{\fbox{Fig. 4}}

Now, let us look at the case of $n\sim 1$ and $J/t\sim 0$.
In the early stage of VMC studies, the stability of the GWF
and various ordered states is discussed.~\cite{YSA,Gros,GL,YSS}
They unanimously concluded a superconducting state with
d$_{x^2-y^2}$-wave symmetry is the most stable in this region.
First, we consider the GWF.
In Fig.4 each component of energy is shown.~\cite{YSA}
A conspicuous aspect different from the 1D case is the
behavior of $E_J/J$ as $n\rightarrow 1$;
$E_J/J\propto (1-n)^{0.7}$, in contrast to $E_t/t$,
$E_3/J_3$, which behave linearly.
This fact means that the GWF is unstable in itself against
the phase separation for $J/t>0$,
because $\partial^2E/\partial n^2<0$.
This instability is intrinsic in the 2D GWF, and
cannot be observed in the stabler states,
as will see below.
\par

Next, we discuss ordered states, namely an AF state
$\Psi_{\rm AF}$: $\eta(r)=1,\ \Phi=\Phi_{\rm AF}$,
and a superconducting state $\Psi_{\rm SC}$:
$\eta(r)=1,\ \Phi=\Phi_{\rm BCS}$.
Here $\Phi_{\rm AF}$ is a Hartree-Fock-type AF wave function,~\cite{YSA}
and $\Phi_{\rm BCS}$ is the BCS-type function,
for which we take different gap symmetries.
In the following, $\Psi_{\rm SC}$ is used by fixing the
electron number,~\cite{Gros,GL}
which agrees with the method of the grand canonical
ensemble~\cite{YSS} in the thermodynamic limit.
\par
\marginpar{\fbox{Fig. 5}}

Figure 5 shows the energy expectation values of these states
for $J/t=0.5$ as a function of $n$.
As mentioned above, the GWF is unstable near $n=1$,
while $\Psi_{\rm AF}$ and $\Psi_{\rm SC}$ of the d$_{x^2-y^2}$
symmetry are not.
Except for the vicinity of the half-filling, d$_{x^2-y^2}$-wave
superconducting state is the most stable.
For s-type and d$_{xy}$ symmetries, no energy reduction
from the GWF is obtained in this parameter regime.
These calculations are consistent with the previous
works.~\cite{YSA,Gros,GL,YSS}
\par

Other symmetries of $\Psi_{\rm SC}$ may be possible.
Actually, we have found that many mixed-symmetry states
including s+$i$d are degenerate with the d$_{x^2-y^2}$-wave
state in the half-filling.
Detailed studies on this problem are needed for $n<1$.~\cite{Joynt}

\par
The $\Psi_{\rm TLL}$ is extended to 2D by Valenti and Gros.~\cite{VG}
They found $\Psi_{\rm TLL}$ shows power-law behaviors for
correlation functions also in 2D.
However, the energy lowering and the critical exponent
obtained are very small, as compared with the 1D system.
Furthermore, since the variational energy of this type
of functions is primarily determined by short-distance
correlation factor,
we can obtain comparable stability by $\Psi_{\rm FL}$.
Thus a different approach may be needed to confirm the
realization of the TLL state in 2D.~\cite{CL2D}
\par

For low electron density, accurate properties are
relatively easy to obtain.
For the supersymmetric case, as in 1D, the GWF is stable
against both $\Psi_{\rm SC}$ and $\Psi_{\rm TLL}$
approximately for $n<0.16$.
However, in 2D this ``free electron" state is
unstable against $\Psi_{\rm SC}$ with d$_{x^2-y^2}$
symmetry for higher density.
\par

Recently Hellberg and Manousakis have obtained
a phase diagram in low electron density by solving
few-body problems.~\cite{HellManou}
They found that in the low density limit there
exists the region $2.0<J/t<3.4367$, where a
singlet-pairing state is the most stable.
Prior to this study Dagotto et al. found
$\Psi_{\rm SC}$ with s-wave symmetry lowers the
energy around this region,~\cite{Dagotto}
which is not the case with the 1D system.
At any rate, the energy lowering by $\Psi_{\rm SC}$
and $\Psi_{\rm TLL}$ is so small that it is not easy to
determine the definite phase boundary in this region,
except for the boundary to the phase separation.
\par

Lastly, we mention the effect of ${\cal H}_3$.
Broadly speaking, this term contributes favorably
to the s-type wave in the low density,
but unfavorably to the d$_{x^2-y^2}$ wave
for every density, when we take the GWF as the standard.
Hence, a ferromagnetic ($J_3<0$) coupling stabilizes
the d$_{x^2-y^2}$-wave superconductivity.
\par

Details of our 2D studies will be published elsewhere.~\cite{YO3}

\section{One-Dimensional Supersymmetric $t$-$J$ Model
with Long-Range Coupling of Inverse Square}

It was shown that the model of the section title is
exactly soluble and the ground state is the GWF.~\cite{KY}
The model is written as
\begin{equation}
{\cal H}=\sum_{i < j}
\Bigl[-t_{ij}\sum_\sigma
      \bigl(c_{i\sigma}^\dagger c_{j\sigma}+{\rm H.c.}\bigr)
     +J_{ij}\bigl({\bf S}_i\cdot{\bf S}_j-\frac{1}{4}n_in_j\bigr)
\Bigr],
\end{equation}
where we require $t_{ij}=J_{ij}/2=t/[(N/\pi)\sin\{\pi$
$\times (x_i-x_j)/N\}]^2\}$, with $t>0$.
It has been discussed that the model has a hidden symmetry
called the Yangian~\cite{htani} which explains the unexpected
degeneracy in the spectrum.~\cite{wlc}
Nevertheless, explicit description of the wave functions
has not been completed.
\par
One can use the VMC technique in judging whether a trial
function is an exact eigenfunction or not by checking
the variances of total energy in VMC sweeps,
as described in \S2.
In this way we have constructed some low-lying
excited states in terms of the Gutzwiller-type
projection of the noninteracting states of up- and down-spin
electrons.~\cite{YK}
Before we present new results, let us explain the notation
and some previous results.

\par
A projected determinantal wave function ${\cal P}\Phi(\{{\bf k}\})$,
which is our trial state, is specified by the occupied
$k$-points in the free electron state $\Phi(\{{\bf k}\})$
from among as many configurations as
$_NC_{N_{\rm \uparrow}}\cdot_NC_{N_{\rm \downarrow}}$,
where $N_\sigma$ denotes the number of electrons with
spin $\sigma$.
In calculations in this section, we use systems of $N=4I+2$
($I$: integer) and $N_\sigma$ being odd integer with
the periodic boundary condition.
For convenience, we divide these functions into two classes:
1) states with continuous occupation of $k$-points and
2) otherwise.
The class 1) has been studied,~\cite{YK} but the class 2)
remains to be investigated.

\par
The states in the class 1) can be described using $N_\sigma$ and current
$J_\sigma$ which is taken to be even integer.
For the state with current $J_\sigma$, we construct $\Phi(\{{\bf k}\})$
in which electrons occupy the states with
$-k_{{\rm F}\sigma}+\pi J_\sigma/N\le k\le k_{{\rm F}\sigma}+\pi J_\sigma/N$.
Here the Fermi wave number $k_{{\rm F}\sigma}$ is given by
$k_{{\rm F}\sigma}=\pi (N_\sigma -1)/N$.
The resultant free-electron state is represented by
$\Phi_{\rm F}(N_\uparrow,J_\uparrow;N_\downarrow,J_\downarrow)$.
Thereby a generalized Gutzwiller wave function is obtained as
$\Psi_{\rm G}(N_\uparrow,J_\uparrow;N_\downarrow,J_\downarrow)=
{\cal P}\Phi_{\rm F}(N_\uparrow,J_\uparrow;N_\downarrow,J_\downarrow)$.
On the other hand, the class 2) consists of a variety of states,
which are generally not easy to be specified concisely.
However, as will be seen below, the eigenstates in this class
are restricted to the states in which only one electron or two are
excited from the states in the class 1).

\par
Let us first review the case in the class 1).
In this class, basic states are those without currents;
$J_\uparrow=J_\downarrow=0$.
In this case $\Psi_{\rm G}$ is an eigenfunction for every
allowed values of the density $n=(N_\uparrow+N_\downarrow)/N$ and
the magnetization $m=(N_\uparrow-N_\downarrow)/N$.
By investigating the dependence of the energy $E_N(m,n)$ on
$N$, $m$ and $n$, it is found that the numerical results are fitted
completely well by the ``experimental" formula:
\begin{equation}
\frac{E_N(m,n)}{t}=
-\frac{\pi^2}{12}(|m|^3-3m^2+n^3-3n^2+4n)
+\frac{\pi^2}{3N^2}(|m|+2n-3),
\label{eq:energy}
\end{equation}
which has now also been obtained analytically.~\cite{HH}
The main term was first obtained by the asymptotic
Bethe Ansatz method.~\cite{KawaABA}
\par
\marginpar{\fbox{Fig. 6}}
Next we consider results for finite currents.
We find that the range of $\Psi_{\rm G}$ being an eigenfunction is
restricted by certain condition for $N_\sigma$ and $J_\sigma$.
In the following, we choose a typical system: $n=0.5$ and $N=60$.
In Figs.6(a) and (b) the circles (filled or empty) show that
$\Psi_{\rm G}(N_\uparrow,J_\uparrow;N_\downarrow,J_\downarrow)$
is an eigenstate at these values of $N_\sigma$ and $J_\sigma$.
The states without circles are found not to be eigenstates.
 From these results we obtain the ``experimental" condition for
$\Psi_{\rm G}(N_\uparrow,J_\uparrow;N_\downarrow,J_\downarrow)$
to be an eigenstate:
\begin{equation}
|J_\sigma|\le N_\sigma+1,
\end{equation}
together with
\begin{equation}
\frac{|J_\uparrow-J_\downarrow|}{2}\le
\frac{|N_\uparrow-N_\downarrow|}{2}+1.
\end{equation}
The critical value of current given by eq.(4.3) corresponds to
a momentum distribution with spin $\sigma$ where the bottom
of the band becomes empty by the momentum shift.
On the other hand eq.(4.4) suggests that a similar band exists
for fictitious particles which are responsible for spin excitations.
An exception to the condition appears in the special
case of $N_\sigma=1$.~\cite{YK}
\par

By calculating the energy increment $\Delta E$ caused by spin
and charge currents, we find that it is fitted excellently
by the formula
\begin{equation}
\frac{\Delta E}{tN}=\Bigl(\frac{\pi}{2N}\Bigr)^2
\bigl[(1-n)J_{\rm c}^2+(1-m)J_{\rm s}^2\bigr],
\label{eq:current}
\end{equation}
where spin and charge currents are given by
$J_{\rm s}=(J_\uparrow-J_\downarrow)/\sqrt 2$ and
$J_{\rm c}=(J_\uparrow+J_\downarrow)/\sqrt 2$, respectively.
The formula applies regardless of values of $N_\sigma$,
and now has also been obtained analytically.~\cite{HH}

\par
Now we proceed to the states in the class 2).
To search eigenstates from a huge number of candidates
in this class, we first check small systems
($N_{\rm e}=6$,10) thoroughly,
then confirm the regularity for larger systems
($N_{\rm e}=30$, etc.).
As a result, we find the following facts.
Eigenstates are necessarily the states in which at most
two electrons are excited from one of the states in the
class 1).
Possible excited states as eigenstates are dependent
on the values of $N_\sigma$ and $J_\sigma$ in the
original state in the class 1).
\par
On account of the limited space, here we mention the cases
except for $N_\sigma=1$~\cite{YK} and $n=1$.~\cite{SH}
In fact, the excluded cases have more abundant
eigenstates probably due to their higher symmetry;
we will report them elsewhere.
In the present cases, allowed states are further restricted
to those with only one excited electron.
\par
Before going to the results, let us see the representation
of excitations appearing in the discussion below.
In these excitations, one electron situated at some $k$-point
in the continuous occupation of the original class-1) state
is excited to the nearest unoccupied $k$-point.
In particular, the electron at the middle of the cluster
may move to both sides.
This type of excitation is indicated by $\ell$ and
spin $\sigma$ as in Fig. 7.
\marginpar{\fbox{Fig. 7}}
\par
First we consider the non-magnetic case.
In Fig. 6(a) possible excitation from each original class-1)
eigenstate (mother state) is summarized for the same system.
The mother states with empty circles have no excited eigenstate
(daughter state) from them, and ones with filled circles have
at least one daughter state.
L-shaped boxes with numbers of $\ell$ show that mother states
in them have daughter states of the specified $\ell$.
A state with an arrow means only the spin specified by
the arrow can excite, and for a state without an
arrow, electrons of either spin direction can excite.
Therefore, for instance, for the mother state
$(J_\uparrow,J_\downarrow)=(-10,-12)$ there exists
a daughter state indicated by [2,$\downarrow$].
For the states with $\ell=-8,-7,7$ and 8, the situations
are a little complex; for the mother state (0,2)
there are two daughter states [8,$\uparrow$] and [$-7,\downarrow$].
Moreover, four daughter states belong to the ground
state (0,0). However these excitations are high energy
processes, so that low-energy particle-hole excitations
from the ground state cannot be described by this type
of wave functions.
\par
Next we take up the cases with finite magnetization.
In Fig.6(b) possible excitations are depicted similarly.
In this case, since available spins for excitation are
common in the rectangular box of $\ell$, we show arrows
together with $\ell$.
The low-energy excitations are prohibited from
the ground state (0,0) also in this case.
\par
To summarize, we have applied the VMC method to identify
considerable amount of exact excited states, which are
described by generalized Gutzwiller functions.
These calculations are a kind of ``experiment".
Part of the eigenfunctions are consistent with the recent
analytic theory.~\cite{KK}
However, some eigenfunctions in the case of $n=1$ are
unexpected.
We hope that the VMC results will stimulate further
analytic theory on this supersymmetric model.
\par

\section{Summary}
In this paper we have described recent developments and new
results for $t$-$J$-type models by the VMC method.
As we have seen above, as well as in refs.\ 7 and 8, the VMC
method have contributed to various issues of the strong
correlation.
Keeping its merits and demerits in mind, one can
develop it further also in other systems of interest, for example,
the multiband models~\cite{SF,AO} and the fractional quantum Hall
effect.~\cite{FQHE}
\par

\centerline{\bf Acknowledgments}
\par\noindent
One of the authors (H.Y.) is grateful to H.\ Shiba for
valuable comments on the manuscript, and to Yusuke Kato for
useful discussions.
This work is partly supported by Grant-in-Aid for Scientific
Research on Priority Areas, ``Computational Physics as a New Frontier in
Condensed Matter Research" and ``Science of High-$T_{\rm c}$
Superconductivity",
from the Ministry of Education, Science and Culture.
\par

\vfil\eject
\noindent{\bf Fig.\ 1}\par\noindent
Comparison of the total energy per site between the GWF
and Bethe ansatz~\cite{Bares} and the non-interacting system
for $J/t=2$ as a function of electron density.
Although the VMC results for $N=102$ is plotted,
the analytic expression is available.~\cite{VMG,YO2}

\noindent{\bf Fig.\ 2}\par\noindent
Phase diagram of the 1D $t$-$J$ model obtained by the
$\Psi_{\rm TLL}$.
The curves show the contours of constant correlation
exponent $K_\rho$.
The used system has 100 sites.

\noindent{\bf Fig.\ 3}\par\noindent
Charge susceptibility in unit of $1/t$ vs.\ $n$
for some values of $J/t$.
Symbols are the results of the $\Psi_{\rm TLL}$.
Solid lines for $J/t=0$ and 2~\cite{KawaTJ} represent the exact
analytic values.
Dashed line is for the non-interacting system.
Dotted lines for $J/t=1.0$ and 2.5 are a guide to the eyes.
The sizes of the symbols represent the relative amplitude of possible
error.
$50\sim 210$-site systems are used.

\noindent{\bf Fig.\ 4}\par\noindent
Expectation values of each energy components
$E_t=\langle{\cal H}_t\rangle$,
$E_J=\langle{\cal H}_J\rangle$,
$E_3=\langle{\cal H}_3\rangle$
as a function of electron density.
The used systems are $N=10\times 10\sim 26\times 26$
with the periodic ($x$ axis) and antiperiodic ($y$ axis)
boundary condition.

\noindent{\bf Fig.\ 5}\par\noindent
Comparison among some variational energies
as a function of electron density.
The used systems are $N=10\times 10$ (diamond),
$12\times 12$ (square), $14\times 14$ (circle),
$16\times 16$ (downward triangle), $20\times 20$
(upward triangle) with the same boundary
condition with Fig.4.

\noindent{\bf Fig.\ 6}\par\noindent
Condition of $N_\sigma$ and $J_\sigma$ for being an eigenstate
in the class 1).
The states specified by solid and open circles are eigenstates.
Two typical cases of different values of $m$ are shown:
(a) non-magnetic case, (b) finite magnetization.
The size of the system and the electron density are fixed at
$N=60$ and $n=0.5$, respectively.
The difference of solid and open circle, boxes, arrows and
numbers indicate possible excited eigenstates in the class 2)
constructed from the original eigenstate in the class 1);
see text and Fig. 7 for detailed explanation.

\noindent{\bf Fig.\ 7}\par\noindent
Representation of excited states (class 2)) from an
original state in the class 1).
Shown is only the $\sigma$-spin configuration, which varies in
the excitation, in $k$-space. The configuration of $-\sigma$-spin
does not change.
An solid (open) circle represents an occupied (unoccupied)
$k$-point.
[$\ell,\sigma$] denotes the following excitation;
the $|\ell|$-th ($1\le|\ell|\le (N_\sigma+1)/2$) electron
(spin $\sigma$) from the nearer boundary,
namely $k={\rm sgn}(\ell)k_{{\rm F}\sigma}+\pi J_\sigma/N$,
is displaced to the nearest unoccupied $k$-point from the
original class-1) state, which is specified by $J_\sigma$
and $N_\sigma$, as in Figs. 6(a) and (b).



\begin{thebibliography}{99}
\bibitem{note15} Recall that for the ground state the variation
principle is always valid, for an excited state, however,
the trial function has to be orthogonal to all the eigenstates
with lower energies.
\bibitem{note3} In cases in which the trial function has small
overlap with different eigenstates from the one aimed at,
the minimum of $\sigma_E$ is often situated near the
energy minimum.~\cite{Gros2}
Nevertheless, the relation (2.5) does not mean that
the minimum of $\sigma_{E}$ agrees with the energy minimum.~\cite{GL}
\end{thebibliography}
\end{document}